\documentclass{article}

\usepackage{arxiv}

\usepackage[utf8]{inputenc} % allow utf-8 input
\usepackage[T1]{fontenc}    % use 8-bit T1 fonts
\usepackage{hyperref}       % hyperlinks
\usepackage{url}            % simple URL typesetting
\usepackage{booktabs}       % professional-quality tables
\usepackage{amsfonts}       % blackboard math symbols
\usepackage{nicefrac}       % compact symbols for 1/2, etc.
\usepackage{microtype}      % microtypography
\usepackage{lipsum}
\usepackage{color}
\usepackage{soul}
\usepackage[version=4]{mhchem}
\usepackage{graphicx}
\usepackage{gensymb}

\title{Molybdenum dichalcogenide cathodes for aluminium-ion batteries}

\author{
  Shalini Divya, James H.\ Johnston\\
  School of Chemical and Physical Sciences\\
  Victoria University of Wellington\\
  Wellington, New Zealand\\
  \And
  Thomas Nann\thanks{Corresponding author.}\\
  School of Mathematical and Physical Sciences\\
  The University of Newcastle\\
  Callaghan, NSW 2308, Australia\\
  \texttt{email:\ \href{mailto:thomas.nann@newcastle.edu.au}{\color{blue}thomas.nann@newcastle.edu.au}}\\
}

\begin{document}
\maketitle

\begin{abstract}
Many successful battery electrodes are based on 2D-layered materials. We have studied aluminium-ion batteries using molybdenum dichalcogenides: \ce{MoS2}, \ce{MoSe2} and MoSSe as active cathode materials. The batteries showed clear discharge voltage plateaus in the ranges 1.6 - 1.4 V for \ce{MoS2} and \ce{MoSe2}, and 0.6 - 0.5 V for MoSSe. \ce{MoS2} and \ce{MoSe2} have similar crystal structures, interestingly we found that \ce{MoSe2} performed better than \ce{MoS2}. MoSSe exhibited a higher specific capacity over \ce{MoS2} and \ce{MoSe2}, but the energy density was lower than \ce{MoSe2} at a current rate of 40 mA g$^{-1}$. \ce{MoSe2} cells recorded a discharge capacity of $\sim$ 110 mAh g$^{-1}$ with an average potential in the range of 2.0 - 1.8 V and 1.5 - 0.8 V during discharge. The cells were stable at 100 mA g$^{-1}$ for over 200 cycles with 90\% coulombic efficiency.
\end{abstract}

% keywords can be removed
\keywords{Aluminium-ion battery \and Layered compounds \and Molybdenum dichalcogenides \and Cathode \and Electrochemistry}

\section{Introduction}
Aluminium-ion batteries (AIBs) offer an alternative to the prevalent lithium-ion battery (LIB) technology. Aluminium being the most abundant metal in the Earth’s crust, these batteries will not only be much cheaper, but hold promise to solve other problems such as recycleability and thermal runaway. Furthermore, the multivalent nature of aluminium may result in a higher specific capacity and energy density compared with other monovalent battery types. Research on AIBs is still in its early stages and current work focuses mainly on electrolytes and cathode materials. In this article, we explore different molybdenum dichalcogenide-based materials and their mechanism of energy storage. We expected that two-dimensional (2D) layered materials that support intercalation of charged species might be suitable as active cathode materials in AIBs \cite{zhang_ultrathin_2015, whittingham_electrical_1976_1}.

The most common AIB electrolyte is currently the ionic liquid 1-ethyl-3-methylimidazolium/tetrachloroaluminate (\ce{[EMIm]+}/\ce{AlCl4-}), although many other alternatives are under investigation \cite{canever_acetamide_2018}. The most studied cathode types are graphite-based, where chloroaluminate ions \ce{AlCl4-} have been shown to intercalate/deintercalate during the charging/discharging processes. Various forms of graphite such as fluorinated graphite \cite{rani_fluorinated_2013}, Kish graphite flakes \cite{wang_kish_2017}, three-dimensional graphitic foam \cite{lin_ultrafast_2015}, few-layer graphene aerogels \cite{qiao_defect-free_2019}, and several others have been tested. Analytical techniques such as X-ray diffraction (XRD), Raman spectroscopy and X-ray photoelectron spectroscopy (XPS) have been used broadly to verify the intercalation/deintercalation mechanism. 

Molybdenum dichalcogenides (\ce{MoX2} where X=S, Se or Te) display similar properties as graphite. They have a 2D-layered structure, which allows intercalation of ions and are electrically conductive. Lower volumetric expansion on cycling is an advantage these materials have over graphitic cathodes \cite{liang_rechargeable_2011, huang_molybdenum_2019}. Amongst various transition metal chalcogenides, \ce{MoS2} has been extensively studied as a cathode for rechargeable batteries \cite{li_mos2_2004, zhu_fast_2015, dong_insights_2019, ding_facile_2012}, making them attractive candidates for AIB cathodes. In 2015, Geng \textit{et al.} found that \ce{Al^3+} ions fully intercalated into chevrel phase \ce{Mo6S8} with the cations occupying two different sites in the crystal lattice \cite{geng_reversible_2015}. This mechanism was called the 'rocking chair' mechanism where charge carrying species shuttled back and forth between intercalating electrodes during cycles while the overall electrolyte concentration remains constant. The discharging and charging reactions at the anode (equation 1) and cathode (equation 2) were proposed as follows:

\begin{align}
          \ce{Al + 7AlCl4^-} &\rightleftharpoons 4\ce{Al2Cl7^- + 3e^-}\\
\ce{8AlCl7^- + 6e^- + Mo6S8 &\rightleftharpoons Al2Mo6S8 + 14AlCl4^-}
\end{align}

Three years later, Li \textit{et al.} prepared \ce{MoS2} microspheres by a simple hydrothermal method \cite{li_rechargeable_2018}. They proposed a similar mechanism where \ce{Al^3+} ions inserted into the electrode accompanied by a phase transformation at the electrode interface. Li and his group confirmed this phase-transition by using \textit{ex-situ} XPS and XRD etching techniques. The reaction equations for this battery system at the cathode (equation 3) and anode (equation 4) were proposed as follows:
\begin{align}
    \ce{MoS2 + x\ce{Al^{3+}}  + 3x\ce{e-} &\rightleftharpoons Al_xMoS2}\\
    \ce{Al} + 7x\ce{AlCl4-} &\rightleftharpoons 4x\ce{Al2Cl7-} + 3x\ce{e-}
\end{align}

In general, these cells showed low energy density and had reversibility issues in the redox processes. It has been reported that transition metal dichalcogenide electrodes tend to transition from a 2H phase into a more conducting 1T phase when used in a battery \cite{fan_hybrid_2017}. A hybrid \ce{Mg^{2+}}/\ce{Li+} cell was tested using bulk \ce{MoS2} as a cathode material. During cyclic voltammetry (CV) scans, the authors associated the first cathodic peak, with a phase transition. 2H phase \ce{MoS2} was converted to 1T phase during initial ion intercalation. This seems to be a common phenomenon for molybdenum dichalcogenides, since Li \textit{et al.} observed similar transitions in sodium ion batteries \cite{li_enhancing_2015, acerce_metallic_2015}. It mostly takes place during the first cycle and since the phase change is irreversible, it can be detected in a cyclic voltammogram.

In this work, we studied a range of 2D molybdenum dichalcogenides including \ce{MoS2}, \ce{MoSe2} and \ce{MoSSe}, and tested them as cathodes for non-aqueous AIBs. 
%This is where I need your comments on DFT calculations...
Our unpublished, preliminary density functional theory (DFT) calculations indicated a significant decrease in inter-layer spacing of these materials when \ce{Al^3+} cations were assumed to intercalate (owing to the very high charge density of \ce{Al^3+}). Therefore, we propose intercalation of structurally distorted \ce{AlCl4-} anions into the cathode layers. Surprisingly we found that \ce{MoSe2}-based cathodes performed different and better than all of the other molybdenum dichalcogenides.

\section{Results and Discussion}
\begin{figure}
  \centering
  \includegraphics[width=0.75\textwidth]{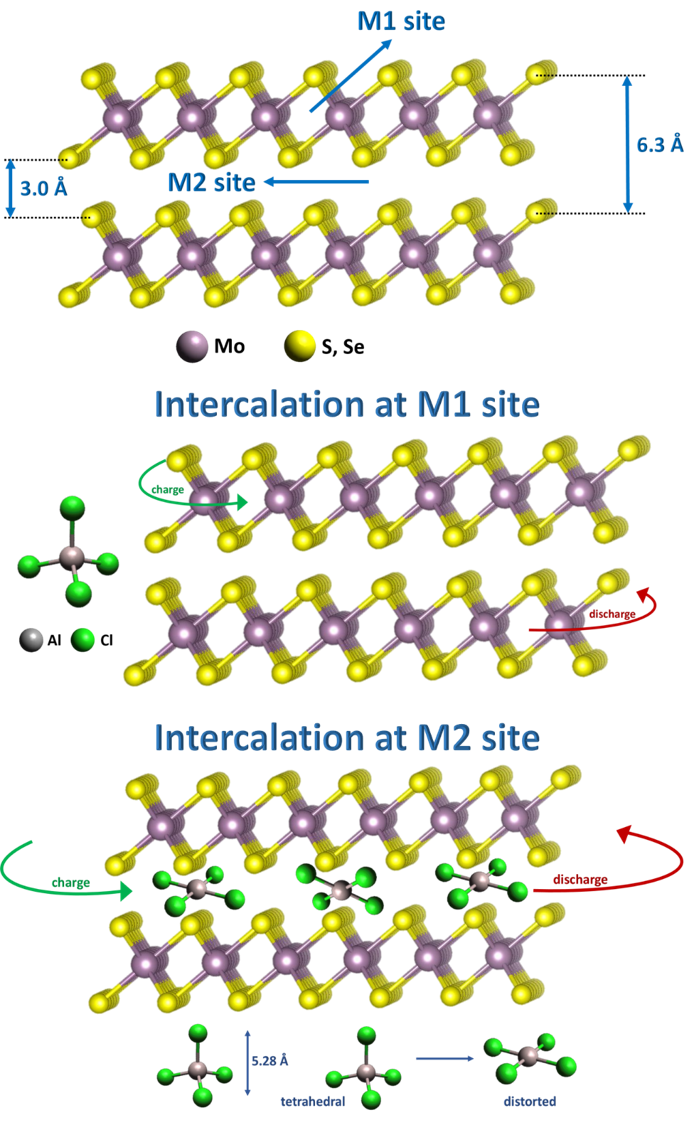}
  \caption{Schematic representation of a) a \ce{MoX2} crystal structure with possible intercalation sites at M1 and M2 b) intercalation at M1 site and c) intercalation at M2 site.}
  \label{figures:S1}
\end{figure}

Figure \ref{figures:S1} shows the crystal structure of \ce{MoX2} where X is sulfur (S) and/or selenium (Se). The material has two vacant sites for intercalation --- M1 and M2. M1 denotes the spaces in between the X-Mo-X atoms, whereas M2 represents the space created between the \ce{MoX2} layers as shown in Figure \ref{figures:S1} a). The inter-layer distance in \ce{MoX2} is 6.3 \AA\ with a gallery height of 3 \AA. The layers are held together by weak van der Waals (vdW) forces. M2 presents an open network and provides various interstitial sites for intercalation. Since \ce{AlCl4-} ions are 5.28 \AA\ in diameter, as reported by Takahashi {\it et al.} \cite{takahashi_niv2o5nh2o_2005}, they undergo some distortion during intercalation to fit into these layers. Our preliminary results showed that \ce{Al^{3+}} would \lq contract\rq\ the \ce{MoX2} layers when trying to intercalate, making \ce{AlCl4-} anion intercalation more likely. Also, the triply charged \ce{Al^{3+}} cation has to overcome strong electrostatic forces from the \ce{S^{2-}} or \ce{Se^{2-}} anion network in order to enter, making the intercalation process slow and most likely not reversible. Therefore, we propose intercalation of \ce{AlCl4-} anions from the electrolyte into M2 sites of \ce{MoX2} during charge. Galvanostatic cycles, cyclic voltammetry (CV), X-Ray diffraction (XRD), Raman spectra and X-Ray Photoelectron Spectroscopy (XPS) results discussed later, strongly support our claim of a reversible intercalation process especially in \ce{MoSe2}.

\begin{figure}
  \centering
  \includegraphics[width=0.4\textwidth]{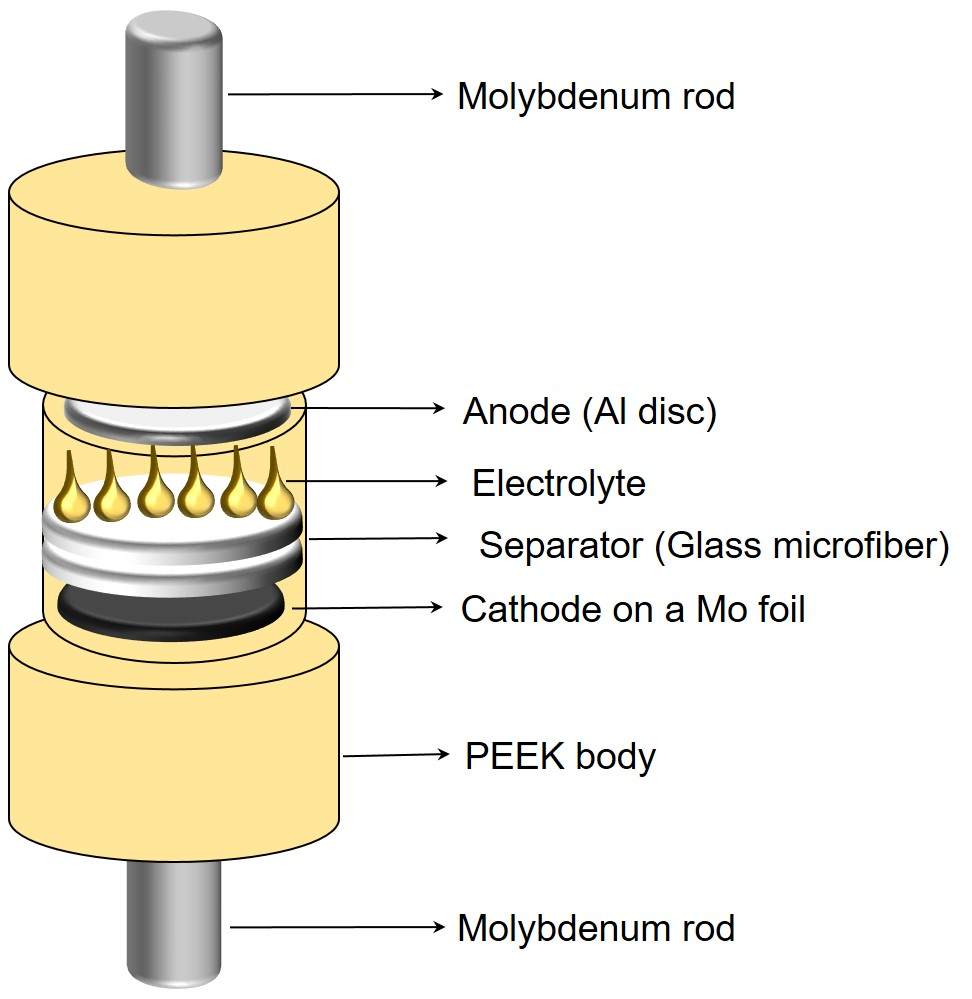}
  \caption{Two-electrode cell assembly using a customised polyether ether ketone (PEEK) body. Molybdenum (Mo) foil was used as the current collector with cathode material coated on top (working electrode) and Mo rods acted as plungers in this Swagelok-type cell. Glass microfibers were used as separators with 99.99\% pure Al foil used as anode.}
  \label{figures:sif1}
\end{figure}

A modified two-electrode polyether ether ketone (PEEK) cell (Figure \ref{figures:sif1}) was used for conducting preliminary electrochemical tests. Cathode preparation, electrolyte synthesis, and cell configuration are described in the Experimental Methods section.

\begin{figure}
  \centering
  \includegraphics[width=\textwidth]{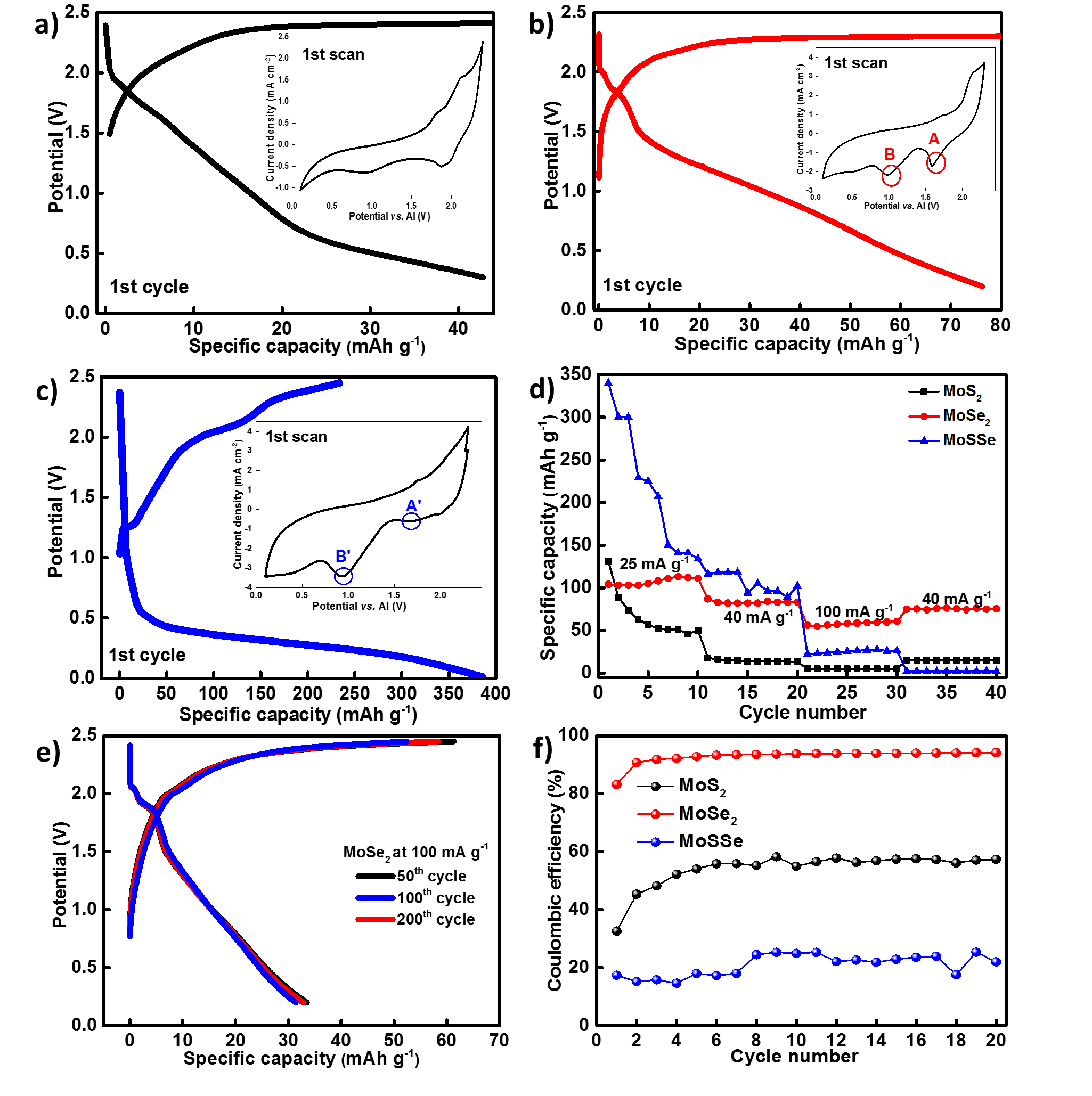}
  \caption{First charge/discharge curve at 40 mA g$^{-1}$ for a) \ce{MoS2}, \textit{inset:}first CV scan of \ce{MoS2}, b) \ce{MoSe2}, \textit{inset:}first CV scan of \ce{MoSe2} and c) MoSSe, \textit{inset:}first CV scan of MoSSe at a scan rate of 10 mV s$^{-1}$ vs.\ Al/\ce{Al^{3+}} electrode. d) Specific capacities of \ce{MoS2}, \ce{MoSe2} and MoSSe at current rates of 25, 40 and 100 mA g$^{-1}$ and then again at 40 mA g$^{-1}$ to test the capacity retention. e) Charge/ discharge performance of \ce{MoSe2} at a high current rate of 100 mA $^{-1}$ for 200 cycles. f) Coulombic efficiencies of \ce{MoS2}, \ce{MoSe2} and MoSSe at a current rate of 100 mA g$^{-1}$ }
  \label{figures:MoX2CDCCV}
\end{figure}

Figure \ref{figures:MoX2CDCCV} a)--c) shows the charge/discharge cycles (CDCs) for \ce{MoS2}, \ce{MoSe2} and MoSSe at a current rate of 40 mA g$^{-1}$. The discharge capacity of \ce{MoS2} in its first cycle was found at $\sim$45 mAh g$^{-1}$, Figure \ref{figures:MoX2CDCCV} a). Comparing this with its first CV scan (Figure \ref{figures:MoX2CDCCV} a \textit{inset}), a good correlation between the discharge voltage plateau and reduction peaks, and other redox features was found. The voltage bend during discharge at 2.0 V matched well with the reduction peak at 2.0 V. The other reduction peak at 1.0 V, however did not correspond to any of the other peaks. With discharge voltage bends between 2.0 - 1.8 V and 1.5 -0.8 V, the first CV scan for \ce{MoSe2} displayed two reduction peaks at 1.65 V (point A) and 1.0 V (point B), Figures \ref{figures:MoX2CDCCV} b) and \ref{figures:MoX2CDCCV} b \textit{inset}). The peak at 1.0 V suggested an irreversible reaction since this peak was absent in the following scans. Based on this, we agree with Li \textit{et al.}'s interpretation and attributed this peak to an irreversible phase transition \cite{li_enhancing_2015}. During this transition, the semi-conducting 2H phase converted into a more metallic 1T phase. This transition seemed to increase the interlayer spacing of \ce{MoSe2} by reducing the vdW forces that exist between the two layers \cite{fan_hybrid_2017}. Al/MoSSe cells showed three distinct plateaus during charge at 1.2 V, 2.0 - 2.1 V and 2.3 - 2.4 V in its first cycle, with a discharge plateau at 0.5 V shown in Figure \ref{figures:MoX2CDCCV} c). Capacities of all molybdenum dichalcogenides were recorded at different current rates of 25, 40 and 100 mA g$^{-1}$, and displayed in Figure \ref{figures:MoX2CDCCV} d). Since \ce{MoSe2} displayed stable specific capacities at all current rates, we recorded further 200 cycles at a high current rate of 100 mA g$^{-1}$. A highly reversible electrochemical reaction was observed since the capacity remained at $\sim$35 mAh g$^{-1}$ after 200 cycles (Figure \ref{figures:MoX2CDCCV} e)) at 100 mA g$^{-1}$. The presence of multiple charging plateaus in MoSSe might correspond to various oxidation processes occurring when \ce{AlCl4-} interacts individually with S and Se atoms. The first CV scan in Figure \ref{figures:MoX2CDCCV} c \textit{inset} showed an irreversible reduction potential at 0.9 V, point B', like \ce{MoSe2}, implying a similar phase transition. It seems MoSSe undergoes a lattice distortion and the material loses its long range order after converting to its 1T phase. This might be the reason why the cells fail to deliver a stable capacity. In addition, a significant difference was observed charge/ discharge curves of all the molybdenum dichalcogenide cathodes-- similar to what was previously reported by Li \textit{et al.} in 2018 \cite{li_rechargeable_2018}. Electrochemical polarisation at the electrode interface might be responsible for the uneven potentials. Formation of an insulating layer on the cathode surface by the new Al-\ce{MoX2} complexes might have attributed to the overpotential.

\begin{figure}
  \centering
  \includegraphics[width=\textwidth]{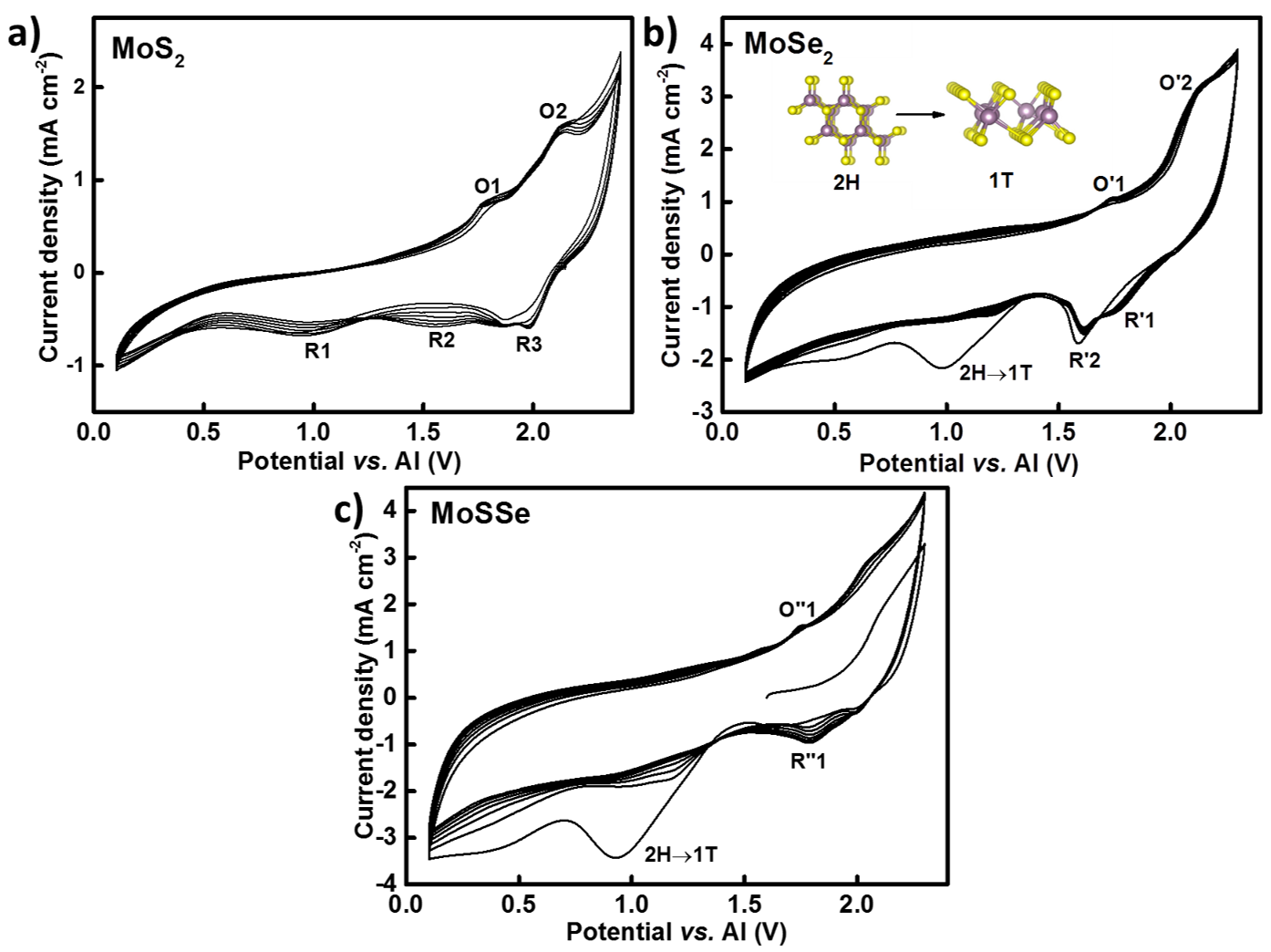}
  \caption{Cyclic voltammograms of a) \ce{MoS2}, b) \ce{MoSe2} and c) MoSSe at a scan rate of 10 mV s$^{-1}$ in a two-electrode aluminium-ion cell against an Al/\ce{Al^{3+}} reference electrode.}
  \label{figures:fig2}
\end{figure}

Electrochemical performance of a blank cell with an uncoated Mo foil (Figure \ref{figures:blankmo}) showed that the current collector did not contribute to the cell's capacity. Both \ce{MoS2} and \ce{MoSe2} have similar interlayer distance (6.3 \AA) and a gallery height of 3.0 \AA. However, \ce{MoSe2} showed a higher capacity and a more stable cycle life. To account for this behaviour, we compared the CVs of all electrodes at a scan rate of 10 mV s$^{-1}$ in Figure \ref{figures:fig2}. Different charge-storage mechanisms lead to distinct features in the CVs. Ideal capacitors result in a rectangular CV shape. Due to the absence of Faradaic processes, the charging/discharging currents become directly proportional to the scan speed. Batteries show oxidation and reduction peaks in their voltammograms because the charge storage takes place via reversible redox processes \cite{jiao_aluminum-ion_2016}. We observed that the CVs of \ce{MoSe2} and MoSSe in Figure \ref{figures:fig2} b) and \ref{figures:fig2} c) covered a broader area suggesting an additional capacitor-like charge storage mechanism. The non-Faradaic process taking place at the surfaces of \ce{MoSe2} and MoSSe might have added to their original capacity values. Also, the peak indicating phase transition from 2H$\rightarrow$1T at $\sim$0.9 - 1.0 V was visible only for \ce{MoSe2} and MoSSe. Hence, it can be concluded that the charge storage in \ce{MoS2} is primarily based on reversible oxidation and reduction of Mo from \ce{Mo^{4+}} to \ce{Mo^{5+}} with oxidation peaks visible at 1.8 V (O1) and 2.1 V (O2), and a corresponding reduction peak at 2.0 V (R3), Figure \ref{figures:fig2} a). Two more reduction peaks were found at 1.6 V (R2) and 0.9 V (R1). However, their peak intensities decreased with every scan. CV scans of Al/\ce{MoSe2} cells in Figure \ref{figures:fig2} b) indicated a reversible electrochemical process, which was in agreement with their CDCs. The scans overlapped with each other displaying two oxidation peaks at 1.7 V (O'1) and 2.1 V (O'2) and corresponding reduction peaks at 1.8 V (R'1) and 1.6 V (R'2). In Figure \ref{figures:fig2} c) an oxidation and a reduction peak at 1.7 V (O''1) and 1.8 V (R''1) was observed for Al/MoSSe respectively. R''1's peak intensity increased after every scan, which might suggest sluggish kinetics in the system; perhaps due to strong interaction between the positive metal ion and the intercalating anion. The voltammogram became more capacitor-like after a few scans, indicating the absence of reversible redox processes. The phenomena indicates that an initial intercalation of anions into the layers of MoSSe was followed by surface adsorption of \ce{AlCl4-} anions.

\begin{figure}
  \centering
  \includegraphics[width=\textwidth]{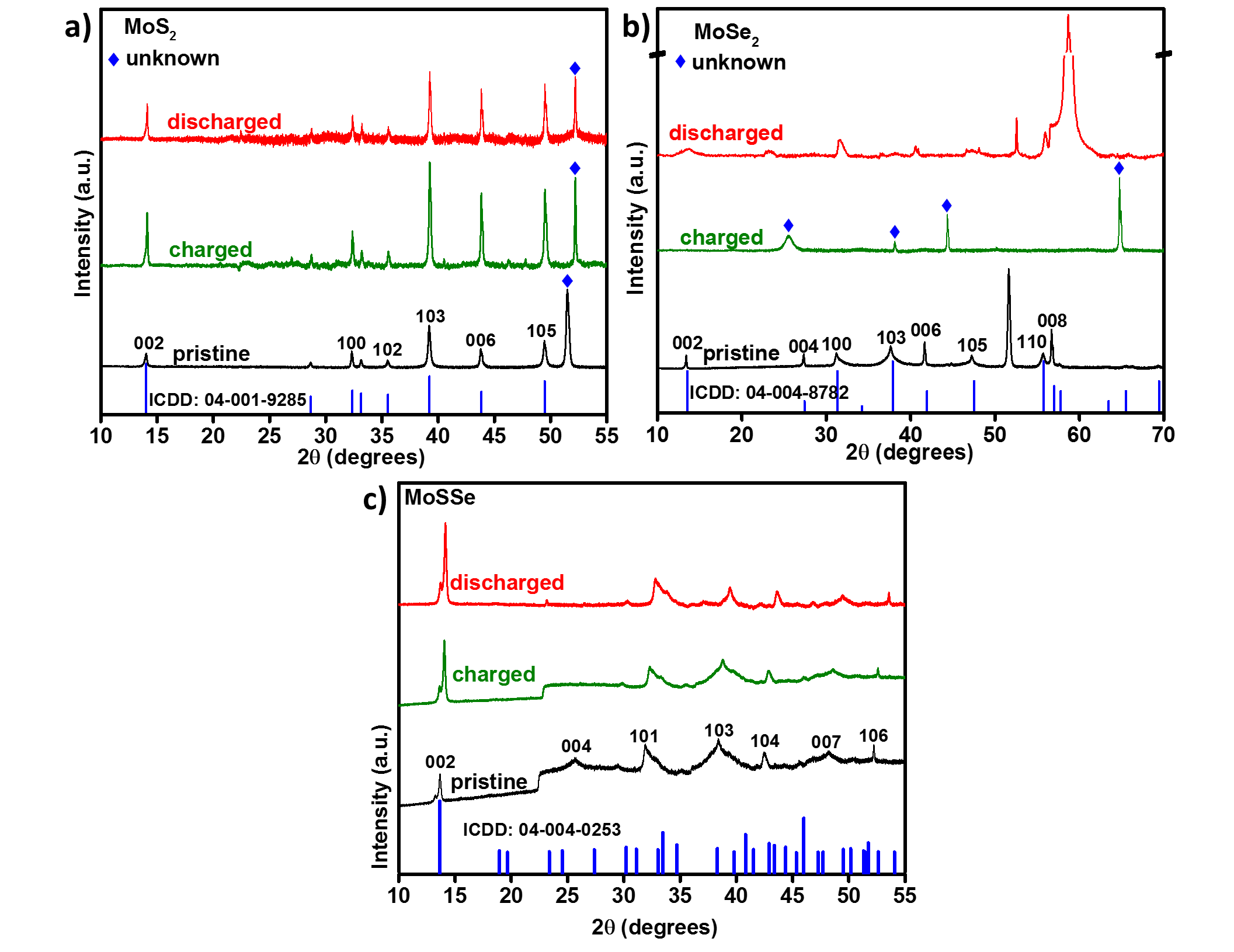}
  \caption{X-ray diffraction patterns of pristine (black), charged (green) and discharged (red) a) \ce{MoS2}, b) \ce{MoSe2} and c) MoSSe electrodes charged to 2.35 V and discharged to 0.2 V vs. Al/\ce{Al^{3+}}, with International Centre for Diffraction Data (ICDD) references, in blue.}
  \label{figures:fig3}
\end{figure}

Figure \ref{figures:fig3} shows the XRD patterns of \ce{MoS2}, \ce{MoSe2} and MoSSe electrodes. Pristine (in black), charged (in green) and discharged (in red) cathodes were compared after 30 cycles each. \ce{MoS2} cells displayed a very small shift in their d-spacings. The peak at 14.21$\degree$ for \textit{002} plane (6.22 \AA) shifted to 14.02$\degree$ (6.32 \AA), as shown in Figure \ref{figures:fig3} a). Most of the peaks retained their positions after charge and discharge showing no significant change in the lattice dimensions. A completely different XRD pattern appeared after charging for Al/\ce{MoSe2} cells, as new peaks appeared at 2$\theta$ values, displayed in Figure \ref{figures:fig3} b). Diffraction peaks of \textit{002}, \textit{100}, \textit{110} and \textit{008} planes reappeared after discharge. Every time the cells were charged, \ce{MoSe2} seemed to adopt this new crystal lattice. However, the characteristic peaks of \ce{MoSe2} reappeared after discharge. This follows closely the observations made by Rani \textit{et al.}, where they proved intercalation of ions into the layers of fluorinated natural graphite during charging using X-ray diffraction data \cite{rani_fluorinated_2013}. This strongly confirms our hypothesis of a reversible intercalation taking place in \ce{MoSe2}. It was interesting to note that MoSSe did not have a well-defined crystal structure to begin with, Figure \ref{figures:fig3} c). The patterns after charge and discharge did not look any different from the untested cathode. This confirmed MoSSe layers did not undergo any significant expansion and the initial specific capacities at $\sim$250 mAh g$^{-1}$ came from the non-Faradaic reactions via electrostatic absorption of the \ce{AlCl4-} anions onto the electrode's surface. Furthermore, the material underwent irreversible changes resulting in cathode degradation after a few cycles. The images obtained from the Scanning Electron Microscope (SEM) have been shown in Figure \ref{figures:sem}. While \ce{MoS2} and \ce{MoSe2} displayed a layered structure in Figure \ref{figures:sem} a) and b) respectively, MoSSe lacked a long-range order. The diffraction patterns of all the Al-Mo-Se complexes present in the ICDD database have been compared with the new complex reported in Figure \ref{figures:fig3}. 

\begin{figure}
  \centering
  \includegraphics[width=\textwidth]{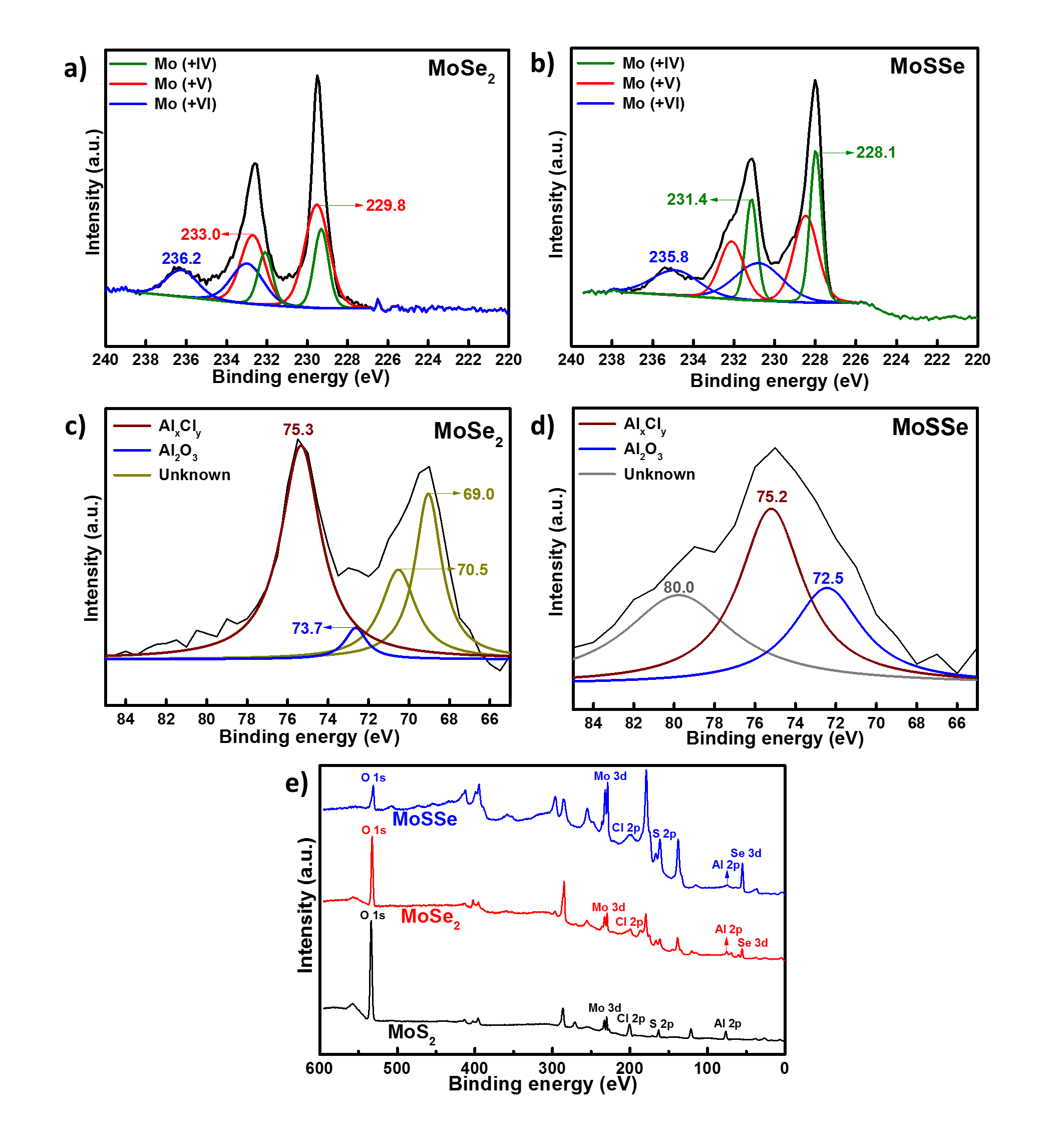}
  \caption{XPS spectra of Mo 3d orbitals in a charged a) \ce{MoSe2} and b) MoSSe cathode and Al 2p orbitals in a charged c) \ce{MoSe2} and d) MoSSe cathode. e) An overview spectrum of all three tested and charged cathodes.}
  \label{figures:MoAlOverallMoSeMoSSe}
\end{figure}

To further understand the interactions between \ce{AlCl4-} and \ce{MoSe2} we used XPS, which is a useful method for distinguishing various oxidation states and helps in identifying different polymorphs (2H and 1T) \cite{fan_hybrid_2017}. The detailed narrow spectrum scans in Figure \ref{figures:MoAlOverallMoSeMoSSe} show the binding energies of Mo (3d$_{5/2}$ and 3d$_{3/2}$, Figure \ref{figures:MoAlOverallMoSeMoSSe} a) and b)) and Al 2p peaks for charged \ce{MoSe2} (Figure \ref{figures:MoAlOverallMoSeMoSSe} c)) and MoSSe electrodes (Figure \ref{figures:MoAlOverallMoSeMoSSe} d)). In pristine \ce{MoSe2}, two peaks appeared at 229.1 eV and 232.2 eV corresponding to 3d$_{5/2}$ and 3d$_{3/2}$ (Figure\ \ref{figures:MoSeSeAlClPrtChg} a)). Selenium displayed a doublet at 55.4 eV and 54.6 eV corresponding to Se 3d$_{3/2}$ and 3d$_{5/2}$ respectively (Figure \ref{figures:MoSeSeAlClPrtChg} c)). Peak splitting in an XPS spectrum can indicate a phase change or a change in oxidation state of the said element. After charge, the peak for Mo 3d split into three doublets indicating the presence of multiple oxidation states or phases of Mo (Figure \ref{figures:MoAlOverallMoSeMoSSe} a)). Se 3d deconvoluted into four peaks after charge, Figure \ref{figures:MoSeSeAlClPrtChg} e), confirming presence of more than one phase after charge. This was similar to observations made by Fan \textit{et al.} where they used \ce{MoS2}/graphene cathode in a hybrid \ce{Mg^2+}/\ce{Li+} cell. Pristine electrodes of MoSSe contained Mo in more than one oxidation state, and provided evidence for the presence of both 2H and 1T polymorphs, Figure\ \ref{figures:MoSeSeAlClPrtChg} b). After charging, the width of peaks at 231.7 eV (Mo 3d$_{5/2}$, in green) and 228.6 eV (Mo 3d$_{3/2}$, in green) increased as displayed in Figure\ \ref{figures:MoAlOverallMoSeMoSSe} b). After comparing Figure\ \ref{figures:MoSeSeAlClPrtChg} d) and f), we noticed that the Se 3d spectrum deconvoluted into four peaks after charging the MoSSe cells. An increase in the peak width was observed for both Mo and Se binding energies. A new peak at $\sim$236 eV in Mo 3d spectra (in blue) for \ce{MoS2}, \ce{MoSe2} and MoSSe electrodes was assigned to Mo$^{6+}$ species typically present in molybdenum oxide, \ce{MoO3}. 

\begin{figure}
  \centering
  \includegraphics[width=\textwidth]{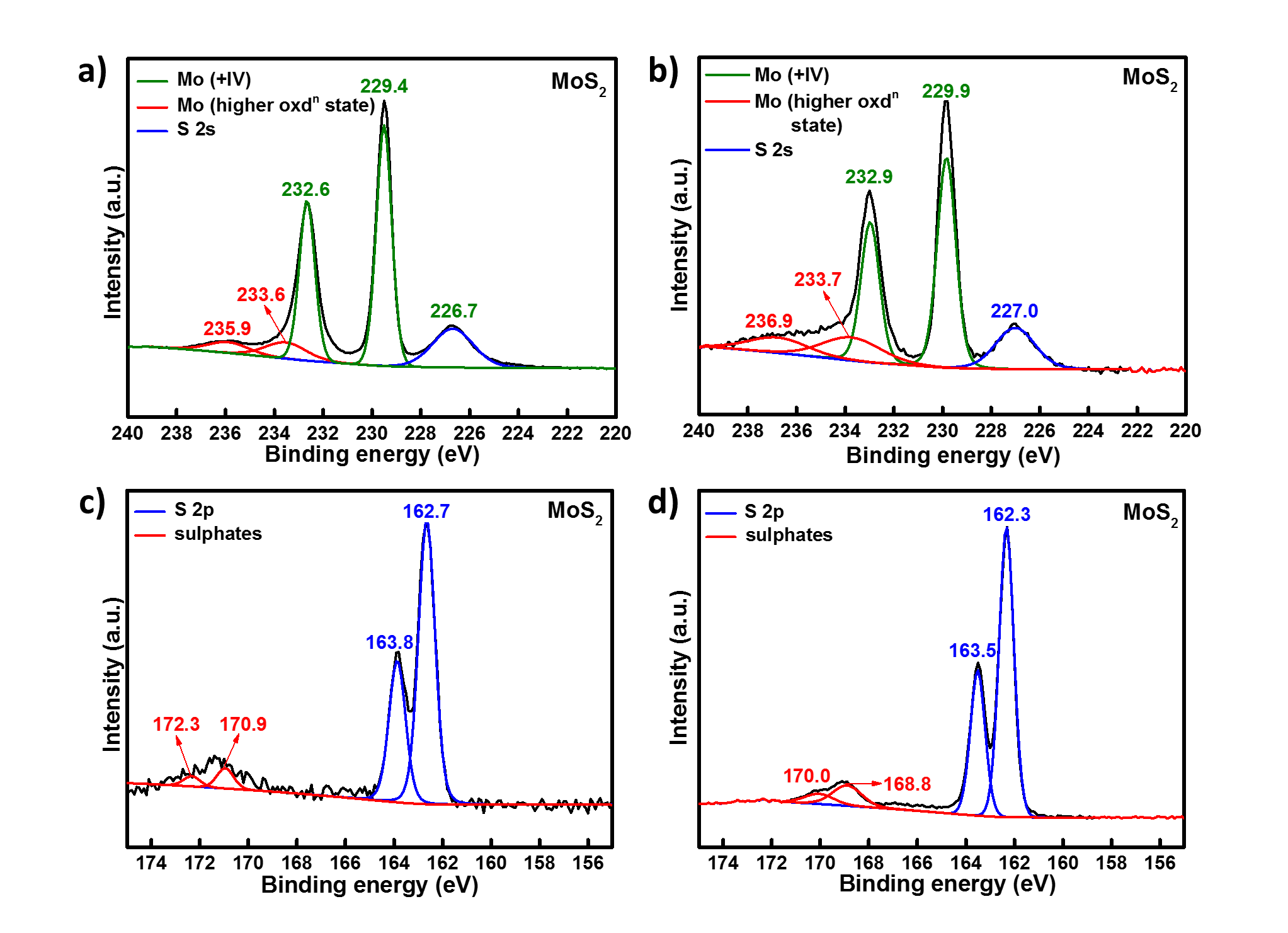}
  \caption{XPS spectra of Mo 3d and S 2s orbital in a charged and b) discharged \ce{MoS2} cathode and binding energies of S 2p orbital in a charged and b) discharged \ce{MoS2} cathode.}
  \label{figures:MoS2XPS}
\end{figure}

\begin{figure}
  \centering
  \includegraphics[width=0.9\textwidth]{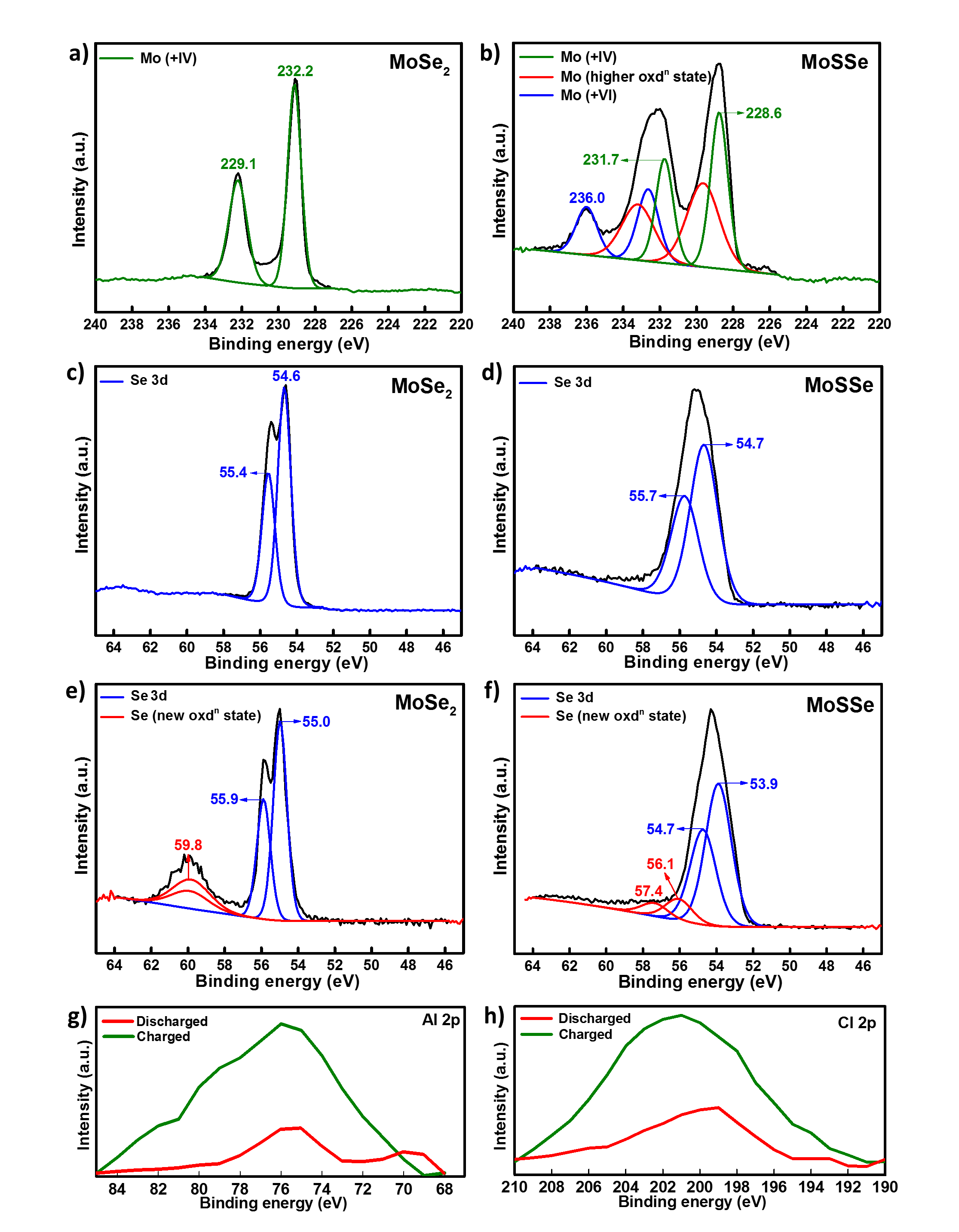}
  \caption{XPS spectra of Mo 3d for pristine a) \ce{MoSe2} and b) MoSSe electrodes. \ce{MoSe2} spectra consists of two peaks at 232.2 eV and 229.1 eV, corresponding to \ce{Mo^{4+}}. MoSSe spectra consists of three doublet bands, which are assigned to \ce{Mo^{4+}}, one with oxidation state between 4 and 6, and another band corresponding to \ce{Mo^{6+}} at 236 eV. Se 3d orbital spectra for pristine c) \ce{MoSe2} and d) MoSSe. Se from \ce{MoSe2} observed peaks corresponding to 3d$_{3/2}$ and 3d$_{5/2}$ at 55.6 eV and 54.6 eV respectively. Binding energies of Se 3d from charged e) \ce{MoSe2} and f) MoSSe cathodes. Binding energies of g) Al 2p and h) Cl 2p in the charged and discharged cathodes-- a general trend followed by all molybdenum dichalcogenides.}
  \label{figures:MoSeSeAlClPrtChg}
\end{figure}

Presence of new peaks and peak shifts that were detected in Mo 3d spectra for charged \ce{MoS2} and \ce{MoSe2} cathodes, were not observed in MoSSe (cf.\ Figures \ref{figures:MoS2XPS} a), b), \ref{figures:MoAlOverallMoSeMoSSe} a), and \ref{figures:MoSeSeAlClPrtChg} a)). This further confirms the absence of redox reactions and that the capacity was mainly derived from a surface-based charge storage. As expected, a general trend was observed for all the three cathodes, where the charged electrodes showed higher concentration of aluminium and chlorine than discharged electrodes as seen in Figure \ref{figures:MoSeSeAlClPrtChg} g) and h). The XPS spectra support the observation that \ce{MoSe2} underwent a phase transformation that made it a better performing cathode than \ce{MoS2}. Further analysis is needed to fully understand the mechanism of MoSSe.

Charged \ce{MoSe2} electrodes displayed binding energies of Al 2p at 77 eV (in red) and 76 eV (in blue) corresponding to chlorides (Al$_x$Cl$_y$) and \ce{Al2O3} respectively in Figure \ref{figures:MoAlOverallMoSeMoSSe} c). New peaks were observed at much lower binding energies --- 69 and 70 eV (in green) suggesting the presence of a new complex with an increased electron density around aluminium. An overall spectra of charged \ce{MoS2}, \ce{MoSe2} and MoSSe cathodes is shown in Figure \ref{figures:MoAlOverallMoSeMoSSe} e) indicating the presence of Al and Cl (from chloroaluminates) and oxygen (from \ce{MoO3}).

\begin{figure}
  \centering
  \includegraphics[width=\textwidth]{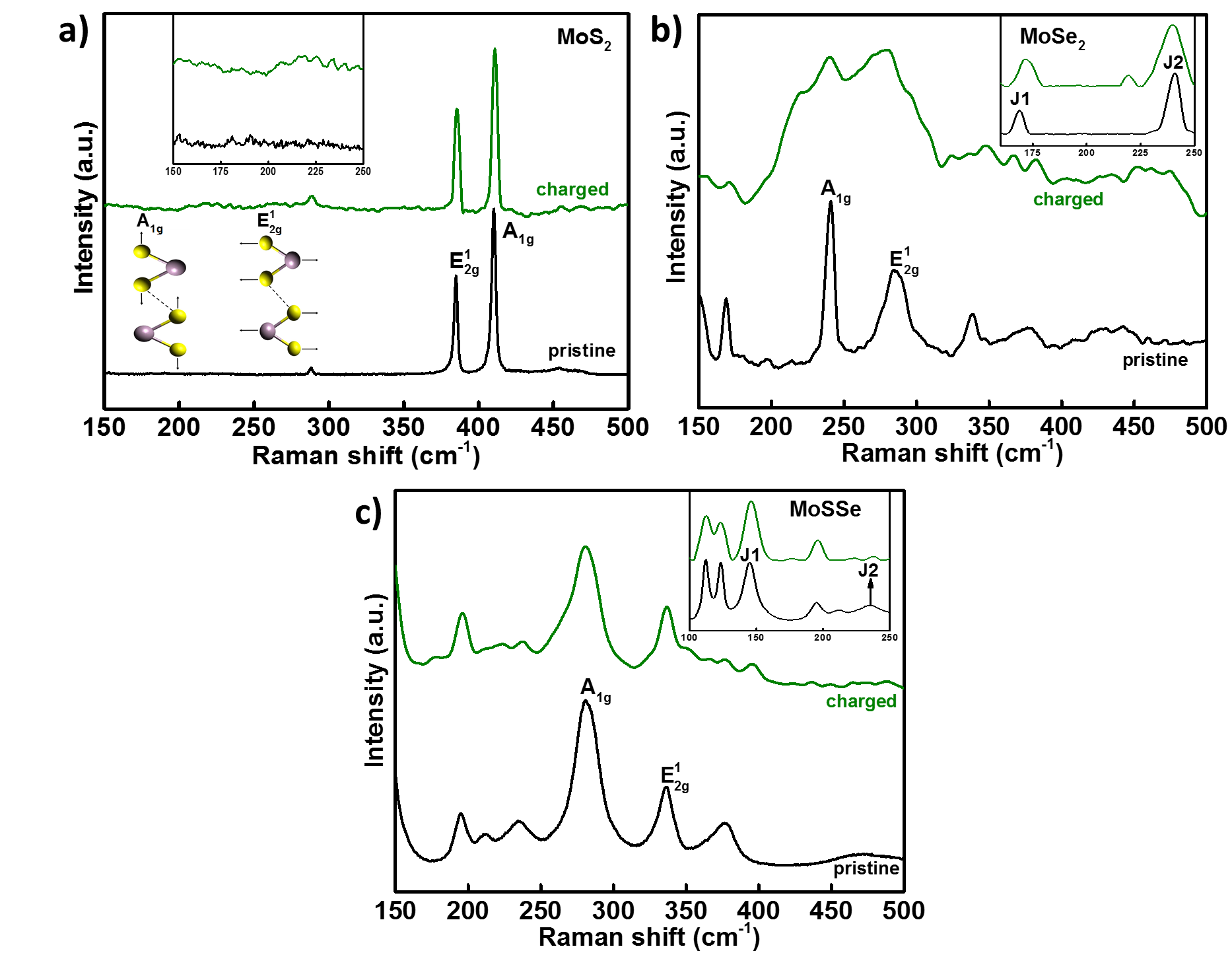}
  \caption{Raman spectra of pristine (black) and charged (green) a) \ce{MoS2}, b) \ce{MoSe2} and c) MoSSe electrodes with position of new Raman active J1 and J2 bands marked along with E$^1_{2g}$ and A$^1_g$ bands.}
  \label{figures/fig5}
\end{figure}

In addition, we compared the Raman spectra of pristine and charged cathodes to detect shifts in vibrational modes shown in Figure \ref{figures/fig5}. Yang \textit{et al.} and Sharma \textit{et al.} have reported that E$^1_{2g}$ and A$^1_g$ are the most intense vibrational modes for molybdenum dichalcogenides \cite{yang_pressure-induced_2019, r_2d_2017,sharma_stable_2018}. Peaks corresponding to E$^1_{2g}$ and A$^1_g$ modes for \ce{MoS2} (Figure \ref{figures/fig5} a)) are prominent at 384.6 cm$^{-1}$ and 410.2 cm$^{-1}$ respectively. A$^1_g$ indicates an out-of-plane symmetric displacement of S atoms, whereas E$^1_{2g}$ suggests an in-layer displacement. Also, separation between the two peaks indicates a multi-layer structure, which was observed for all three materials. No significant peak shift or peak broadening was observed for the charged \ce{MoS2} electrode. For 2H \ce{MoSe2} (Figure \ref{figures/fig5} b)), A$^1_g$ is the most intense vibration occurring at a frequency lower than that of E$^1_{2g}$. When the number of layers decreases, the A$^1_g$ mode softens and an increase in full-width-at-half-maximum (FWHM) is detected. Spectra generated after intercalation were different from the pristine cathodes because phase conversion from 2H to 1T decreases the molecule's symmetry and more Raman bands get active. The presence of J1 and J2 peaks in addition to E$^1_{2g}$ and A$^1_g$ at lower wavelengths suggest the existence of 1T phase especially for \ce{MoSe2} and MoSSe (inset, Figure\ \ref{figures/fig5} b) and c)). This agrees with the CV scans and XPS results where a phase transition was observed for \ce{MoSe2} and MoSSe. Raman results suggest that the symmetry and vibrational modes of \ce{MoSe2}'s crystal lattice changed after repeated cycles. It seems that the phase transformation occurring in the first few cycles for \ce{MoSe2} changes its structure in a way that allows it to intercalate more \ce{AlCl4-} anions resulting in a higher capacity than \ce{MoS2} despite both of them having a similar stricture. First principle studies on both materials shall confirm this hypothesis. 

\section{Conclusions}
In this work, we studied systematically the charging/discharging mechanism of aluminium-ion batteries, using different molybdenum dichalcogenide cathodes. It was found that \ce{MoSe2} showed a higher capacity and cyclic stability than \ce{MoS2} and MoSSe. CV and XPS results indicated an irreversible phase transition to a more metallic 1T phase. This transition worked in favour of \ce{MoSe2} and its capacity increased. XRD, XPS and Raman results supported the hypothesis that \ce{AlCl4-} intercalated reversibly into \ce{MoSe2}. An additional electro-capacitive behaviour was observed in \ce{MoSe2} that added to the its overall capacity. The cells delivered a potential of $\sim$2.0 V with discharge capacity of 30 mAh g$^{-1}$ with nearly 95\%\ coulombic efficiency at a current rate of 100 mA g$^{-1}$.

\section{Experimental Section}
\subsection{Cathode preparation}
A slurry was prepared by mixing MoX$_2$ (85\%\ by wt.), 9\%\  binder (PVDF, MTI Corporation) and 6\%\ Super-P conductive carbon (99+\%\ metals basis, Alfa Aesar) in N-methyl pyrrolidone (NMP) (anhydrous, 99.5\%, Sigma-Aldrich). this slurry was ‘doctor-bladed’ onto molybdenum foil (thickness 0.1 mm, MTI Corporation) and dried in a vacuum oven at 120\degree C for 12 hours to adhere the slurry on the conductive substrate and evaporate the solvent. The specific loading of the active materials was approx.\ 12 mg cm$^{-2}$. 

\subsection{Electrolyte preparation}
Anhydrous AlCl$_3$ (Sigma-Aldrich) and EMImCl (97\%, Sigma-Aldrich) were mixed in a molar ratio of 1.3:1, at room temperature. EMImCl was baked in vacuum for 24 hours at 100\degree C to remove residual moisture. Small aliquots of AlCl$_3$ were added to EMImCl after every few minutes until the white fumes settled down. The ionic liquid was stirred for 2--3 hours until a clear brown liquid was obtained. Since the electrolyte was hygroscopic in nature, it was prepared in a N$_2$-filled glove box with <0.1 ppm \ce{H2O}/\ce{O2}. 

\subsection{Cell assembly}
 Polyether ether ketone (PEEK) cells were used for preliminary electrochemical tests. Molybdenum rods were used as plungers to push in the electrodes as close to each other. Active material coated on molybdenum foil was used as the cathode and placed at bottom of the cell. Since this was a two-electrode setup, aluminium foil was used as both counter and reference electrode. We used Mo foil because nickel and steel showed reactivity towards the ionic liquid electrolyte and reduced its potential window. Glass microfibers (Grade GF/F, Whatman) were used as separators. 80 $\mu$l of the electrolyte were used to wet the separator. Aluminium foil (thickness 0.1 mm, 99\%, GoodFellow) was used as an anode and placed on top of the separator. The cell was then sealed and wrapped with a paraffin to avoid any further air or moisture contact after it was taken out of the glove-box.

\bibliographystyle{unsrt}  
%\bibliography{smp}

\begin{thebibliography}{10}

\bibitem{canever_acetamide_2018}
Nicol{\`o} Canever, Nicolas Bertrand, and Thomas Nann.
\newblock Acetamide: A low-cost alternative to alkyl imidazolium chlorides for
  aluminium-ion batteries.
\newblock {\em Chem. Commun.}, September 2018.

\bibitem{rani_fluorinated_2013}
J.~Vatsala Rani, V.~Kanakaiah, Tulshiram Dadmal, M.~Srinivasa Rao, and
  S.~Bhavanarushi.
\newblock Fluorinated {{Natural Graphite Cathode}} for {{Rechargeable Ionic
  Liquid Based Aluminum}}\textendash{{Ion Battery}}.
\newblock {\em Journal of The Electrochemical Society}, 160(10):A1781--A1784,
  January 2013.

\bibitem{wang_kish_2017}
Shutao Wang, Kostiantyn~V. Kravchyk, Frank Krumeich, and Maksym~V. Kovalenko.
\newblock Kish {{Graphite Flakes}} as a {{Cathode Material}} for an {{Aluminum
  Chloride}}-{{Graphite Battery}}.
\newblock {\em ACS Appl Mater Interfaces}, 9(34):28478--28485, August 2017.

\bibitem{lin_ultrafast_2015}
Meng-Chang Lin, Ming Gong, Bingan Lu, Yingpeng Wu, Di-Yan Wang, Mingyun Guan,
  Michael Angell, Changxin Chen, Jiang Yang, Bing-Joe Hwang, and Hongjie Dai.
\newblock An ultrafast rechargeable aluminium-ion battery.
\newblock {\em Nature}, 520(7547):324--328, April 2015.

\bibitem{qiao_defect-free_2019}
Jia Qiao, Haitao Zhou, Zhongsheng Liu, Hejing Wen, and Jianhong Yang.
\newblock Defect-free soft carbon as cathode material for {{Al}}-ion batteries.
\newblock {\em Ionics}, February 2019.

\bibitem{liang_rechargeable_2011}
Yanliang Liang, Rujun Feng, Siqi Yang, Hua Ma, Jing Liang, and Jun Chen.
\newblock Rechargeable {{Mg Batteries}} with {{Graphene}}-like {{MoS2 Cathode}}
  and {{Ultrasmall Mg Nanoparticle Anode}}.
\newblock {\em Adv. Mater.}, 23(5):640--643, 2011.

\bibitem{huang_molybdenum_2019}
Junda Huang, Zengxi Wei, Jiaqin Liao, Wei Ni, Caiyun Wang, and Jianmin Ma.
\newblock Molybdenum and tungsten chalcogenides for lithium/sodium-ion
  batteries: {{Beyond MoS2}}.
\newblock {\em Journal of Energy Chemistry}, 33:100--124, June 2019.

\bibitem{li_mos2_2004}
Xiao-Lin Li and Ya-Dong Li.
\newblock {{MoS2 Nanostructures}}:\, {{Synthesis}} and {{Electrochemical Mg2}}+
  {{Intercalation}}.
\newblock {\em J. Phys. Chem. B}, 108(37):13893--13900, September 2004.

\bibitem{zhu_fast_2015}
Changbao Zhu, Xiaoke Mu, Peter~A. {van Aken}, Joachim Maier, and Yan Yu.
\newblock Fast {{Li Storage}} in {{MoS2}}-{{Graphene}}-{{Carbon Nanotube
  Nanocomposites}}: {{Advantageous Functional Integration}} of {{0D}}, {{1D}},
  and {{2D Nanostructures}}.
\newblock {\em Adv. Energy Mater.}, 5(4):n/a--n/a, February 2015.

\bibitem{geng_reversible_2015}
Linxiao Geng, Guocheng Lv, Xuebing Xing, and Juchen Guo.
\newblock Reversible {{Electrochemical Intercalation}} of {{Aluminum}} in
  {{Mo6S8}}.
\newblock {\em Chem. Mater.}, 27(14):4926--4929, July 2015.

\bibitem{li_rechargeable_2018}
Zhanyu Li, Bangbang Niu, Jian Liu, Jianling Li, and Feiyu Kang.
\newblock Rechargeable {{Aluminum}}-{{Ion Battery Based}} on {{MoS2 Microsphere
  Cathode}}.
\newblock {\em ACS Appl. Mater. Interfaces}, 10(11):9451--9459, March 2018.

\bibitem{fan_hybrid_2017}
Xin Fan, Rohit~Ranganathan Gaddam, Nanjundan~Ashok Kumar, and Xiu~Song Zhao.
\newblock A {{Hybrid Mg2}}+/{{Li}}+ {{Battery Based}} on
  {{Interlayer}}-{{Expanded MoS2}}/{{Graphene Cathode}}.
\newblock {\em Adv. Energy Mater.}, 7(19), October 2017.

\bibitem{li_enhancing_2015}
Yifei Li, Yanliang Liang, Francisco~C. Robles~Hernandez, Hyun Deog~Yoo, Qinyou
  An, and Yan Yao.
\newblock Enhancing sodium-ion battery performance with interlayer-expanded
  {{MoS2}}\textendash{{PEO}} nanocomposites.
\newblock {\em Nano Energy}, 15:453--461, July 2015.

\bibitem{takahashi_niv2o5nh2o_2005}
Katsunori Takahashi, Ying Wang, and Guozhong Cao.
\newblock {{Ni}}-{{V2O5}}{$\cdot$}{{nH2O Core}}-{{Shell Nanocable Arrays}} for
  {{Enhanced Electrochemical Intercalation}}.
\newblock {\em J. Phys. Chem. B}, 109(1):48--51, January 2005.

\bibitem{jiao_aluminum-ion_2016}
Handong Jiao, Junxiang Wang, Jiguo Tu, Haiping Lei, and Shuqiang Jiao.
\newblock Aluminum-{{Ion Asymmetric Supercapacitor Incorporating Carbon
  Nanotubes}} and an {{Ionic Liquid Electrolyte}}:
  {{Al}}/{{AlCl3}}-[{{EMIm}}]{{Cl}}/{{CNTs}}.
\newblock {\em Energy Technol.}, 4(9):1112--1118, September 2016.

\bibitem{yang_pressure-induced_2019}
Linfei Yang, Lidong Dai, Heping Li, Haiying Hu, Kaixiang Liu, Chang Pu, Meiling
  Hong, and Pengfei Liu.
\newblock Pressure-induced metallization in {{MoSe}} 2 under different pressure
  conditions.
\newblock {\em RSC Adv.}, 9(10):5794--5803, 2019.

\bibitem{r_2d_2017}
Rao C.~N. R and Waghmare~Umesh Vasudeo.
\newblock {\em 2d {{Inorganic Materials Beyond Graphene}}}.
\newblock {World Scientific}, August 2017.

\bibitem{sharma_stable_2018}
Chithra~H. Sharma, Ananthu~P. Surendran, Abin Varghese, and Madhu Thalakulam.
\newblock Stable and scalable {{1T MoS2}} with low temperature-coefficient of
  resistance.
\newblock {\em Sci Rep}, 8, August 2018.

\bibitem{acerce_metallic_2015}
Muharrem Acerce, Damien Voiry, and Manish Chhowalla.
\newblock Metallic {1T} phase {MoS} 2 nanosheets as supercapacitor electrode materials
\newblock{\em Nature Nanotech}, 10(4):313-318, April 2015.

\bibitem{ding_facile_2012}
Shujiang Ding, Dongyang Zhang, Jun Song Chen, and Xiong Wen (David) Lou.
\newblock Facile synthesis of hierarchical MoS 2 microsphere.
\newblock{\em Nanoscale}, 4(1):95-98, 2012.

\bibitem{dong_insights_2019}
Yulian Dong, Yang Xu, Wei Li, Qun Fu, Minghong Wu, Eberhard Manske, Jorg Kroger, and Yong Lei.
\newblock Insights into the Crystallinity of Layer-Structured Transition Metal Dichalcogenides on Potassium Ion Battery Performance: A Case Study of Molybdenum Disulfide.
\newblock{\em Small}, 15(15):1900497, 2019.


\bibitem{zhang_ultrathin_2015}
Hua Zhang
\newblock Ultrathin Two-Dimensional Nanomaterials
\newblock{\em ACS Nano}, 9(10):9451-9469, October 2015.

\bibitem{whittingham_electrical_1976_1}
M. S. Whittingham
\newblock Electrical {Energy} {Storage} and {Intercalation} {Chemistry}.
\newblock{\em Science}, 192(4244):1126-1127, June 1976.

\bibitem{li_cobalt_doped_2018_1}
Bing Li, Yuxin Shi, Kesheng Huang, Mingming Zhao, Jiaqing Qiu, Huaiguo Xue, and Huan Pang.
\newblock Cobalt-Doped Nickel Phosphite for High Performance of Electrochemical Energy Storage.
\newblock{\em Small}, 14(13):1703811, 2018.

\bibitem{li_ultrathin_2017_1}
Bing Li, Peng Gu, Yongcheng Feng, Guangxun Zhang, Kesheng Huang, Huaiguo Xue, and Huan Pang.
\newblock Ultrathin Nickel–Cobalt Phosphate 2D Nanosheets for Electrochemical Energy Storage under Aqueous/Solid-State Electrolyte.
\newblock{\em Adv. Funct. Mater.}, 27(12):1605784, 2017.

\bibitem{xiao_synthesis_2020_1}
Xiao Xiao, Lianli Zou, Huan Pang, and Qiang Xu.
\newblock Synthesis of micro/nanoscaled metal–organic frameworks and their direct electrochemical applications.
\newblock{\em Chem. Soc. Rev}, 49(1):301-331, January 2020.

\bibitem{li_ultrathin_2018_1}
Bing Li, Peng Gu, Guangxun Zhang, Yao Lu, Kesheng Huang, Huaiguo Xue, and Huan Pang.
\newblock Ultrathin Nanosheet Assembled Sn0.91Co0.19S2 Nanocages with Exposed (100) Facets for High-Performance Lithium-Ion Batteries.
\newblock{\em Small}, 14(5):1702184, 2018.

\end{thebibliography}

\clearpage
\newpage

%%%%%% Supporting Information %%%%%%%

\section*{Supporting Information}

\begin{center}
  \textbf{\Huge Molybdenum dichalcogenide cathodes for aluminium-ion batteries}\vspace{0.5cm}
  
  \textit{Shalini Divya, James H.\ Johnston, Thomas Nann*}
\end{center}

\begin{figure}[h!]
  \centering
  \includegraphics[width=\textwidth]{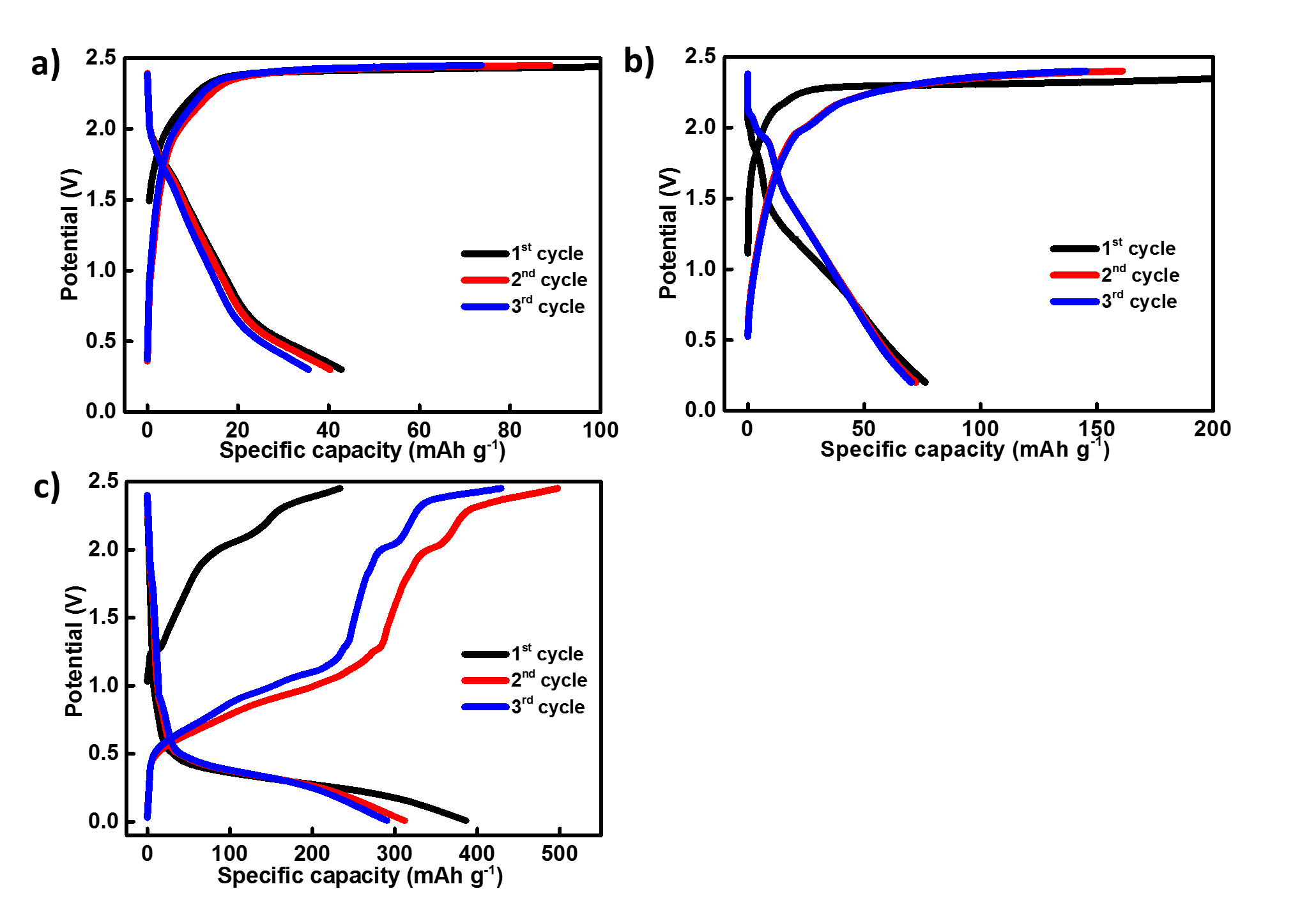}
  \caption{Galvanostatic charge/discharge profile of the first three cycles of a) \ce{MoS2}, b) \ce{MoSe2} and c) MoSSe at a current rate of 40 mA g$^{-1}$ in a two-electrode setup.}
  \label{SF:MoX23cycles}
\end{figure}

\begin{figure}[h!]
  \centering
  \includegraphics[width=\textwidth]{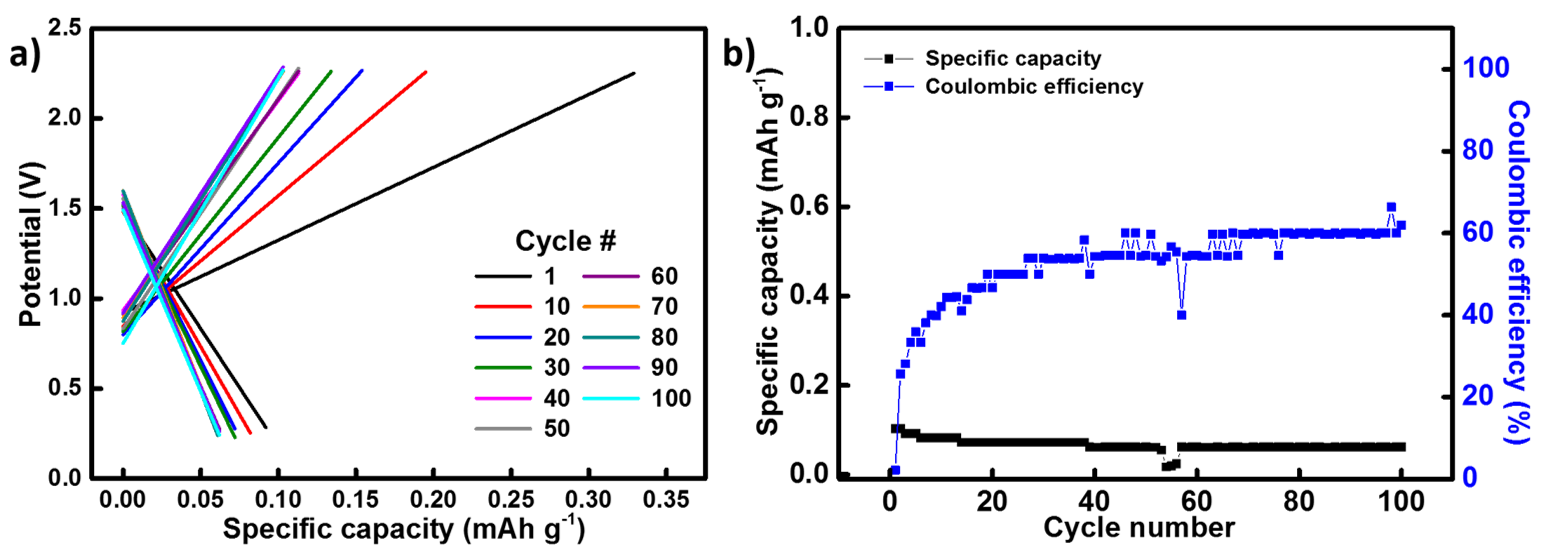}
  \caption{a) Galvanostatic charge/discharge curves of an Al/ blank Mo cell using a two-electrode setup at a current rate of 40 mA g$^{-1}$. The cells failed to achieve any significant specific capacities during both charge and discharge. b) Blank Mo foil displayed CE at 40\%. This confirmed that when acting as the current collector, molybdenum did not contribute any capacity of its own..}
  \label{figures:blankmo}
\end{figure}

\begin{figure}[h!]
  \centering
  \includegraphics[width=\textwidth]{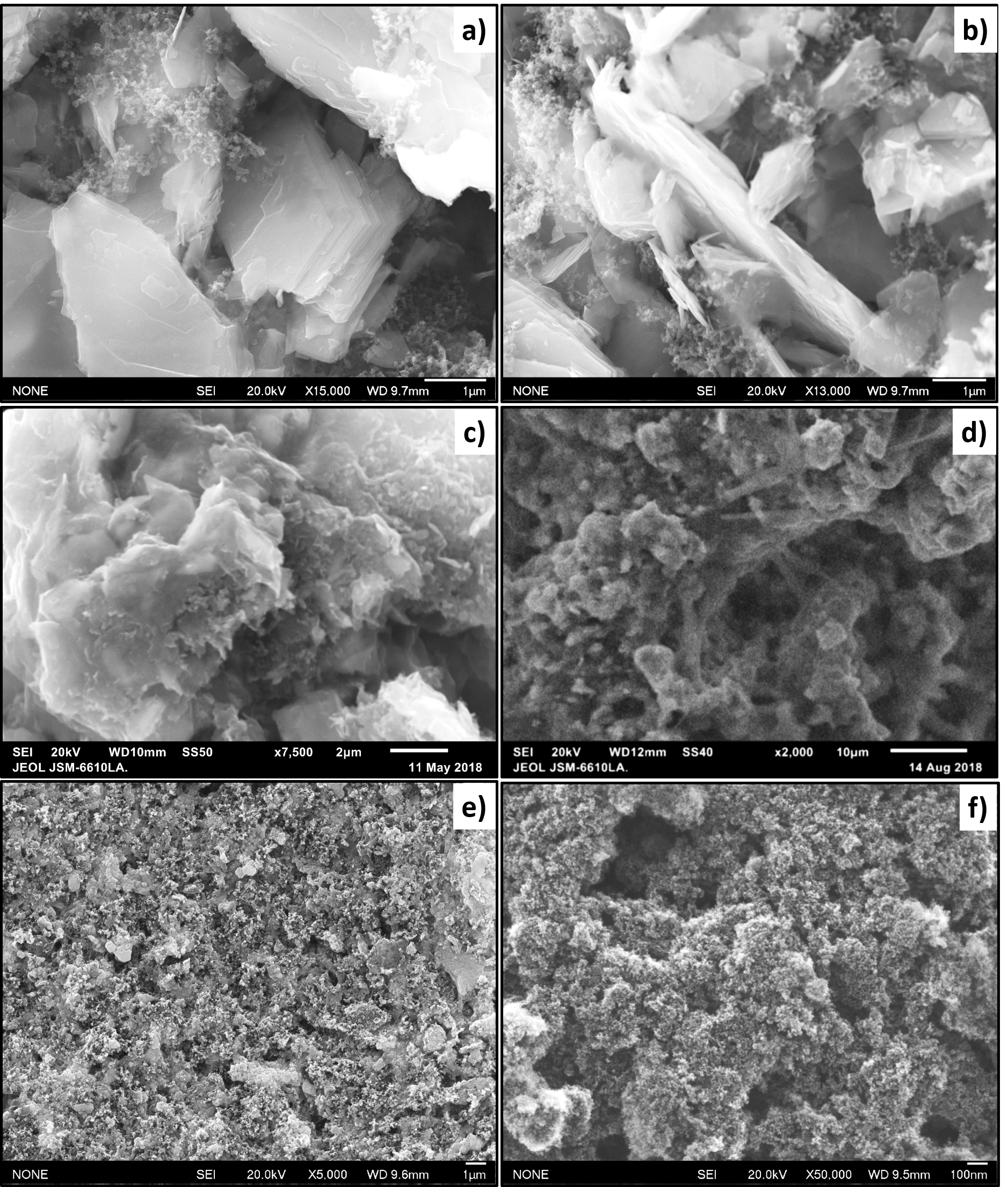}
  \caption{SEM images of pristine a) \ce{MoS2} and b) \ce{MoSe2}; and cycled c) \ce{MoS2} and d) \ce{MoSe2}. SEM images of e) pristine and f) cycled MoSSe. \ce{MoS2} and \ce{MoSe2} clearly have a layered structure, while MoSSe lacks a long-range order.}
  \label{figures:sem}
\end{figure}

\begin{figure}[h!]
  \centering
  \includegraphics[width=\textwidth]{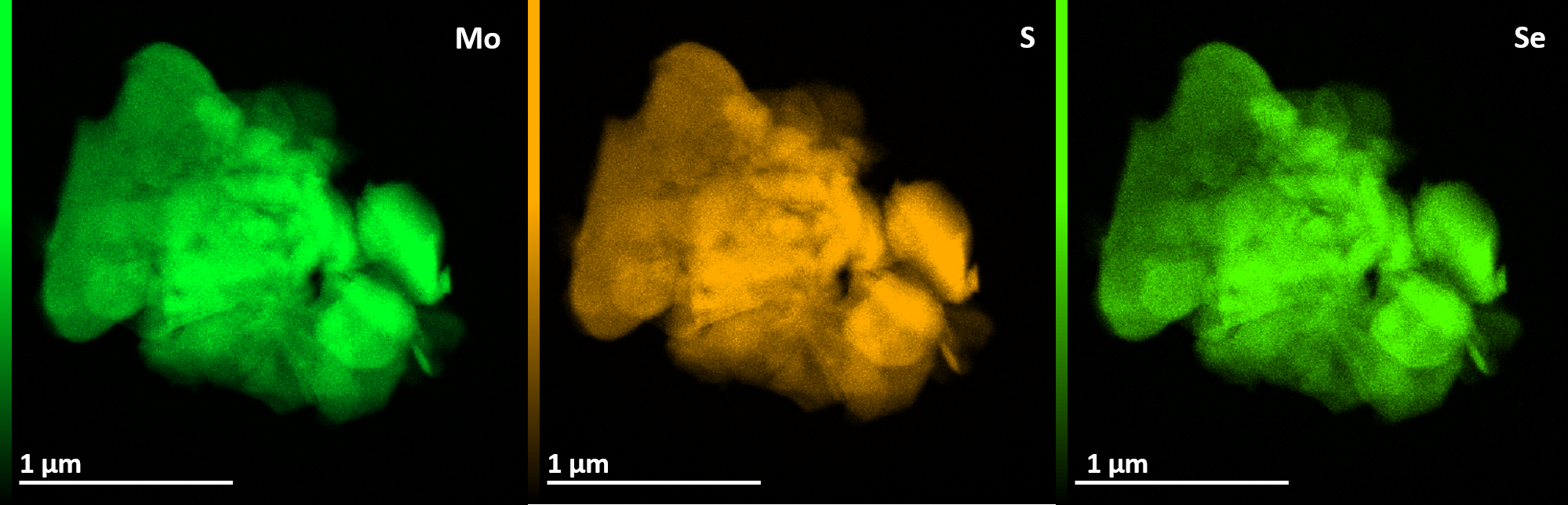}
  \caption{Energy dispersive X-ray spectroscopy (EDXS) map of pristine MoSSe showing equal distribution of Mo, S and Se.}
  \label{figures:MoSSeEDS}
\end{figure}

\begin{figure}[h!]
  \centering
  \includegraphics[width=\textwidth]{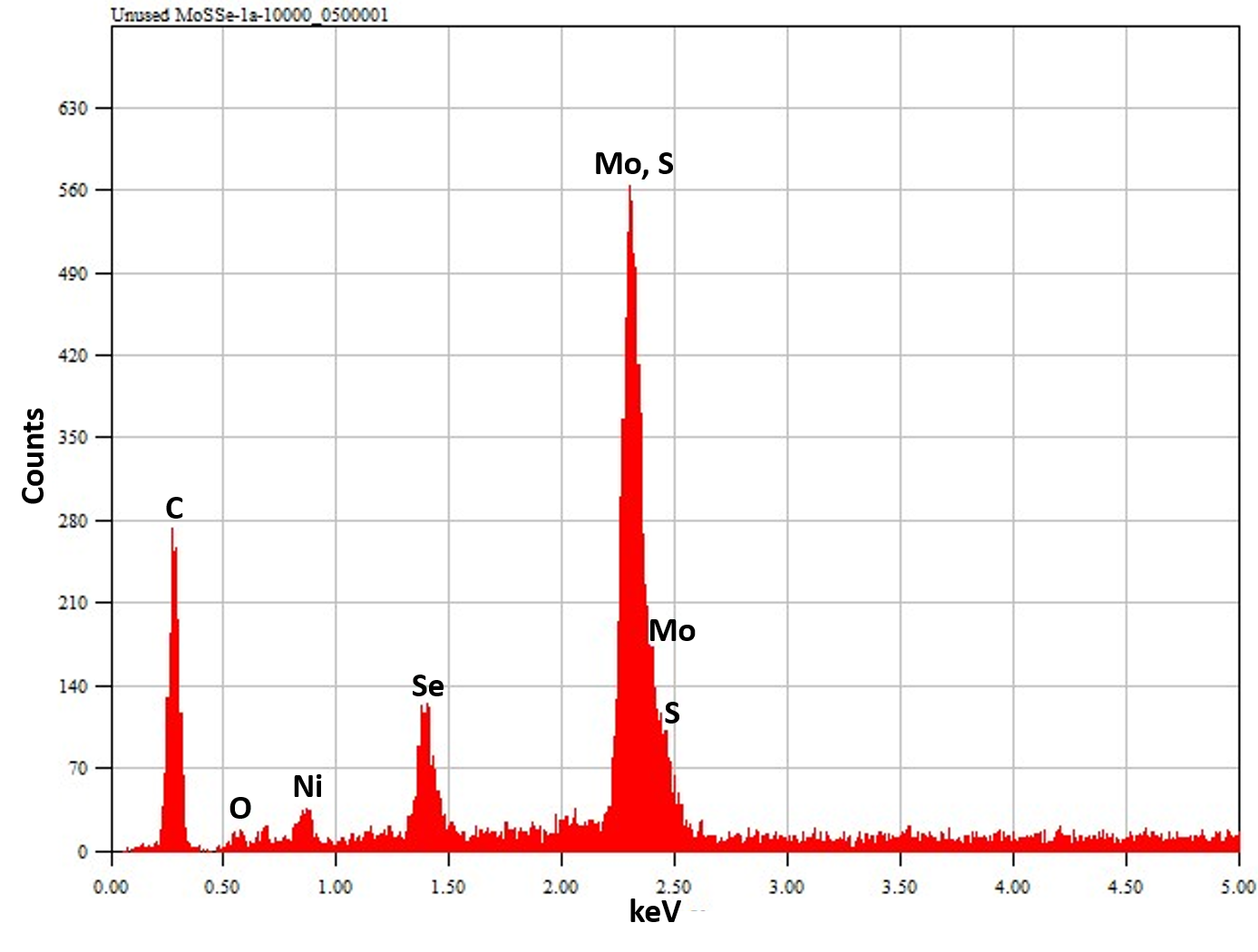}
  \caption{EDX spectra of pristine MoSSe showing peaks for Mo, S and Se.}
  \label{figures:mosseprt}
\end{figure}

\end{document}